\documentclass[12pt,a4paper,sort&compress]{article}

\pdfoutput=1

\usepackage{tikz}
    \usetikzlibrary{positioning}
    \usetikzlibrary{arrows}
    \usetikzlibrary{shapes}
    \usepackage{pgfplots}
    \usetikzlibrary{calc}
    \usetikzlibrary{decorations.markings}
     \usetikzlibrary{decorations.pathmorphing}
    \tikzset{snake it/.style={decorate, decoration=snake}}
\def\centerarc[#1](#2)(#3:#4:#5) 
    { \draw[#1] ($(#2)+({#5*cos(#3)},{#5*sin(#3)})$) arc (#3:#4:#5); }
    \usepackage{booktabs}
\usepackage{jheppub}
\usepackage{amsmath}
\usepackage{amssymb}
\usepackage{amsfonts}
\usepackage{amsthm}
\usepackage{shuffle}
\usepackage{enumitem}
\usepackage{color}   
\usepackage{hyperref}
\usepackage{bbm}
\usepackage{multirow}

\newcommand{\ord}{\begin{cal}O\end{cal}}

\newcommand{\rd}{\mathrm{d}}

\def\beq{\begin{equation}}
\def\eeq{\end{equation}}
\def\bsp#1\esp{\begin{split}#1\end{split}}

\newenvironment{sloppyequation}[0]{\sloppy\begin{flushleft}\hspace*{0.75cm}\(}{\)\end{flushleft}\fussy}

\newcommand{\beqsloppy}{\begin{sloppyequation}}
\newcommand{\eeqsloppy}{\end{sloppyequation}}

\newcommand{\cD}{\begin{cal}D\end{cal}}

\newcommand{\cI}{\begin{cal}I\end{cal}}

\newcommand{\cR}{\begin{cal}R\end{cal}}

\usepackage{hyperref}
\theoremstyle{definition}

\newcommand{\uz}{{z}}

\newcommand{\uPi}{{\Pi}}
\newcommand{\uPhi}{{\Phi}}
\newcommand{\ut}{{t}}

\DeclareMathOperator{\GL}{GL}
\DeclareMathOperator{\SO}{SO}
\DeclareMathOperator{\OO}{O}

\DeclareMathOperator{\NS}{NS}
\DeclareMathOperator{\T}{T}

\DeclareMathOperator{\SL}{SL}
\DeclareMathOperator{\K}{K}

\DeclareMathOperator{\diag}{diag}

\DeclareMathOperator{\ZZ}{\mathbb{Z}}

\allowdisplaybreaks

\title{Feynman integrals, elliptic integrals and two-parameter K3 surfaces}

\author[a]{Claude Duhr}
\emailAdd{cduhr@uni-bonn.de}
\author[a]{Sara Maggio}
\emailAdd{smaggio@uni-bonn.de}
\affiliation[a]{Bethe Center for Theoretical Physics, Universität Bonn, D-53115, Germany
}

\abstract{The three-loop banana integral with three equal masses and the conformal two-loop five-point traintrack integral in two dimensions are related to a two-parameter family of K3 surfaces. We compute the corresponding periods and the mirror map, and we show that they can be expressed in terms of ordinary modular forms and functions. In particular, we find that the maximal cuts of the three-loop banana integral with three equal masses can be written as a product of two copies of the maximal cuts of the two-loop equal-mass sunrise integral. Our computation reveals a hidden symmetry of the banana integral not manifest from the Feynman integral representation, which corresponds to exchanging the two copies of the sunrise elliptic curve.}

\begin{document}

\preprint{BONN-TH-2025-05}

\maketitle


\section{Introduction}
\label{sec:intro}

It is well known that multi-loop Feynman integrals are periods~\cite{MR1852188,Bogner:2007mn}, and they are tightly connected to topics in modern algebraic geometry and number theory. A lot of progress has been made over the last decade in understanding the types of geometries and functions that arise from Feynman integrals (see, e.g., ref.~\cite{Bourjaily:2022bwx} for a recent review). The development of novel methods to compute Feynman integrals and to understand their properties is therefore tightly linked to progress in understanding the associated geometries. 

A lot of information about an algebraic variety is encoded in its periods, obtained by integrating a differential form that defines a non-trivial cohomology class over a cycle. The periods of the geometry attached to a Feynman integral typically compute the maximal cuts of the integral. The latter play an important role in physics. For example, they give valuable information about the class of functions to which the full Feynman integral evaluates, and they also play an important role in many modern approaches to finding so-called systems of canonical differential equations satisfied by the integrals~\cite{Henn:2013pwa,Pogel:2022ken,Pogel:2022vat,Pogel:2022yat,Gorges:2023zgv,Dlapa:2022wdu}. In some cases, the periods are even sufficient to compute the full Feynman integral~\cite{Duhr:2022pch,Duhr:2023eld,Duhr:2024hjf}. 
Computing the periods of an algebraic variety, however, is typically a complicated task, because they usually define new transcendental functions, and one often resorts to numerical techniques for their evaluation (cf.,~e.g.,~ref.~\cite{Lairez:2023nih}). A notable exception are periods of hyperelliptic curves of genus $g$, which can be expressed in terms of Lauricella functions depending on $2g-1$ variables. In the special case $g=1$ of elliptic curves, the periods can be written as complete elliptic integrals of the first kind.
A lot of techniques to compute periods have been developed in the context of Calabi-Yau (CY) varieties. 
Using techniques inspired from CY geometries, it was possible to evaluate the maximal cuts of multi-loop banana integrals~\cite{Bloch:2014qca,MR3780269,Klemm:2019dbm,Bonisch:2020qmm,Bonisch:2021yfw}, ice cone integrals~\cite{Duhr:2022dxb}, traintrack and fishnet integrals in two dimensions~\cite{Duhr:2022pch,Duhr:2023eld,Duhr:2024hjf}. CY periods also enter the computation of multi-loop corrections to the photon propagator in Quantum Electrodynamics~\cite{Forner:2024ojj} and certain observables relevant to gravitational wave physics~\cite{Frellesvig:2023bbf,Frellesvig:2024rea,Frellesvig:2024zph,Klemm:2024wtd,Driesse:2024feo,Dlapa:2024cje}.

It is an interesting question if and when periods of one variety can be related to periods of other geometries. This question is not only an interesting area of research in mathematics (cf.,~e.g.,~refs.~\cite{Ruppert:1990aa,Almkvist3,BognerCY,BognerThesis,10.1145/3597066.3597111,Clingher2010LatticePK,doranclingher1,Nagano:2024aa} for examples), but various instances of such relations between different geometries have recently been observed in the context of Feynman integrals. For example, the periods of families of CY threefolds depending on a small number of moduli can be written as periods of hyperelliptic curves~\cite{Jockers:2024uan}. Moreover, it was observed that the genus of the hyperelliptic curve attached to a maximal cut may not be unique, resulting in different expressions for the periods~\cite{Marzucca:2023gto}. Furthermore, the celebrated Basso-Dixon formula~\cite{Basso:2017jwq,Basso:2021omx,derkachov_basso-dixon_2019}, which expresses so-called fishnet integrals as determinants of ladder graphs, can be interpreted in two dimensions as a relation between periods of varieties of different dimensions~\cite{Duhr:2023eld,Duhr:2024hjf}. Finally, it is known that the periods of one-parameter families of CY twofolds -- also known as K3 surfaces -- can be written as products of complete elliptic integrals, i.e., periods of a family of elliptic curves~\cite{doran,BognerCY,BognerThesis}, a result which has been used in a variety of Feynman integral computations~\cite{Bloch:2014qca,MR3780269,Broedel:2019kmn,Broedel:2021zij,Pogel:2022yat,Klemm:2024wtd}. In ref.~\cite{Dlapa:2024cje} also an example of a two-parameter family of K3 surfaces was presented whose periods can be written as a product of periods of two different families of elliptic curves.

The aim of this paper is to present new examples where we can express the periods of a family of K3 surfaces attached to a Feynman integral in terms of a product of elliptic integrals for different elliptic curves. More specifically, we study the family of K3 surfaces attached to the three-loop banana integral with three equal masses and the conformal two-loop five-point traintrack integral in two dimensions. Our result is based on a recent analysis of the modular properties of families of K3 surfaces based on the structure of the transcendental lattice~\cite{lattice_paper}. The monodromy group of a two-parameter family of K3 surfaces is a subgroup of the orthogonal group $\OO(2,2)$, and it is well known that there is a two-to-one map from $\SL(2,\mathbb{R})\times\SL(2,\mathbb{R})$ to (the connected component containing the identity of) $\OO(2,2)$. Depending on the form of the intersection pairing on the transcendental lattice, this group-theoretic argument implies that the periods (and also the mirror map) can be cast in terms of modular forms. We explicitly determine these modular expressions for both the traintrack and banana integrals. For the banana integral we find that the maximal cut of the three-loop banana integral with three equal masses is simply a product of two copies of the maximal cut of the well known equal-mass sunrise integral~\cite{Laporta:2004rb}. Our result for this maximal cut exhibits an additional symmetry, corresponding to exchanging the two copies of the sunrise elliptic curve, which is not manifest at the level of the Feynman integral.

Our paper is organised as follows: In section~\ref{sec:CYs} we give a very brief review of K3 surfaces and their periods, in particular when they admit a parametrisation in terms of modular forms. In sections~\ref{sec:traintrack} and~\ref{sec:bananas} we present the main results of this paper, namely the explicit results for the periods and the mirror map for the traintrack and banana integrals in terms of modular forms and complete elliptic integrals. In ref.~\ref{sec:conclusions} we present our conclusions. We include two appendices where we review results about modular forms used in the main text, and we also discuss a modular parametrisation for the product of elliptic integrals from ref.~\cite{Dlapa:2024cje}.


\section{K3 surfaces and their periods}
\label{sec:CYs}

The main topic of this paper are Feynman integrals related to certain two-parameter families of K3 surfaces. We therefore start by giving a short review of K3 surfaces and their cohomology groups and periods.

A K3 surface is a two-dimensional complex K\"ahler manifold $M$ with vanishing first Chern class. We will consider families of K3 surfaces depending on $m=2$ parameters $z=(z_1,z_2)$. The cohomology groups of $M$ carry a canonical Hodge structure. We will be interested in the middle cohomology, which admits the decomposition
\beq
H^2(M,\mathbb{C}) = H^{2,0}(M)\oplus H^{1,1}(M)\oplus H^{0,2}(M)\,,
\eeq
where $H^{p,q}(M)$ is the space of all cohomology classes of $(p,q)$-forms on $M$, i.e., the space of all differential forms involving exactly $p$ holomorphic and $q$ antiholomophic differentials. The Betti numbers are $h^{p,q} = \dim H^{p,q}(M)$. 
The vanishing of the first Chern class implies $h^{2,0}=h^{0,2}=1$, and so there is a unique 
holomorphic $(2,0)$-form $\Omega$. 
The \emph{periods} of $M$ are defined by integrating $\Omega$ over two-dimensional cycles $H_2(M,\mathbb{Z})$. 

The second cohomology group $H^2(M,\mathbb{Z})$ of a K3 surface is very well understood. In particular, it is always 22-dimensional, and the intersection pairing in homology induces a pairing in cohomology
\beq
Q:H^2(M,\mathbb{Z})\times H^2(M,\mathbb{Z})\to \ZZ\,.
\eeq
The form of the pairing is extremely constrained. It has signature $(3,19)$ and its Gram matrix in a particular basis is completely fixed. In the following, the explicit form of the pairing on the full cohomology group will not be relevant, and only a part of the cohomology group will play a role, as we now explain.

The middle cohomology group admits a direct sum decomposition
\beq
H^2(M,\mathbb{Z}) = \NS(M)\oplus \T(M)\,,
\eeq
where $\NS(M)=H^{1,1}(M)\cap H^2(M,\ZZ)$ is the N\'eron-S\'everi group
and $\T(M) = \NS(M)^{\bot}$ is the transcendental lattice. The N\'eron-S\'everi group can also be characterised as the dual space of all those cycles $\Gamma\in H_2(M,\ZZ)$ such that $\int_{\Gamma}\Omega=0$. Since we are interested in understanding the periods, the N\'eron-S\'everi group does not play any role in the following, and we can focus exclusively on the transcendental lattice $\T(M)$.

The intersection pairing induces a pairing on $\T(M)$. One can show that $\T(M)$ always has dimension $m+2=4$ and the signature of the pairing is $(2,m) = (2,2)$. We can collect the non-vanishing periods into a vector
\beq\label{eq:per_vec_def}
\uPi(z) = \big(\Pi_{0}(z),\Pi_{1}(z),\Pi_{2}(z),\Pi_{3}(z)\big)^T = \Big(\int_{\Gamma_{0}}\!\!\Omega,\ldots,\int_{\Gamma_{3}}\!\!\Omega\Big)^T\,,
\eeq
where the $\Gamma_0,\ldots,\Gamma_{3}$ are linearly independent cycles from $H_2(M,\ZZ)$.
The intersection pairing implies quadratic relations among the periods, known as \emph{Hodge-Riemann bilinear relations},
\beq\bsp
\label{eq:Hodge-Riemann}\uPi(\uz)^T\Sigma\uPi(\uz) &\,= 0\textrm{~~~and~~~}\uPi(\uz)^{\dagger}\Sigma\uPi(\uz) > 0\,.
\esp\eeq
The matrix $\Sigma$ is related to the Gram matrix of the intersection pairing between cycles,
\beq
\big(\Sigma^{-1}\big)^T =  \big(\Gamma_i\cap\Gamma_j\big)_{0\le i,j\le 3}\,.
\eeq
One can show that for K3 surfaces, $\Sigma$ is symmetric and its entries are integers.

Rather than performing the integrations in eq.~\eqref{eq:per_vec_def}, it is typically easier to compute the periods as solutions to some differential equations. These differential equations generate an ideal, called the \emph{Picard-Fuchs differential ideal}.
The periods are not single-valued functions of $\uz$, but they develop a non-trivial monodromy as $\uz$ is varied along a closed loop $\gamma$ encircling one of the singular divisors in the moduli space. The monodromy group $G_{\!M}$ acts linearly via
\beq
\uPi(\uz) \to M_{\gamma}\uPi(\uz)\,,\qquad M_{\gamma}\in G_{\!M}\subseteq \GL(4,\mathbb{Z})\,.
\eeq
The intersection pairing is monodromy-invariant, which puts strong constraints on the possible form of the monodromy group $G_{\!M}$. In particular, $G_{\!M}$ must be a subgroup of the orthogonal group 
\beq
\OO(\Sigma,\ZZ) = \big\{M\in \GL(4,\ZZ): M^T\Sigma M=\Sigma\big\}\,.
\eeq

The analytic structure of the periods is most easily described in the case when the moduli space of complex structure deformations has a point of \emph{maximal unipotent monodromy} (MUM). From now on we assume that our family of K3 surfaces has a MUM-point, and without loss of generality, we may assume that the MUM-point is at $z=0$. Then there is a basis in which 
$\Pi_0(z)$ is holomorphic at $z=0$ and $\Pi_{3}(z)$ diverges as $\log z_k\,\log z_l$. The remaining two period diverge as $\log z_k$ close to the MUM-point. In this basis, the intersection pairing takes the form
\beq
\Sigma = \left(\begin{smallmatrix} 0 & 0 & 1\\ 0&S&0 \\ 1&0&0\end{smallmatrix}\right)\,,
\eeq
where $S$ is a symmetric integer matrix.
Equation~\eqref{eq:Hodge-Riemann} implies that we can fix $\Pi_3(z)$ in terms of the other periods,
\beq
\Pi_3 (z) = -\frac{1}{2\Pi_0(z)}\,\Pi^{(1)}(z)^TS\Pi^{(1)}(z)\,,
\eeq
where we defined $\Pi^{(1)}(z) = \big(\Pi_1(z),\Pi_2(z)\big)^T$.
We normalise the remaining periods according to
\beq
\label{eq:MUM-basis}
\Pi_0(\uz) = 1+\ord(z_i)\,,\qquad \Pi_k(\uz) = \Pi_0(\uz)\,\frac{\log z_k}{2\pi i} + \ord(z_l)\,,\qquad k=1,2\,.
\eeq
We can then define canonical coordinates $q_k = \exp(2\pi i t_k) = z_k+\ord(z_l^2)$ on the moduli space by
\beq\label{eq:t_def}
t_k(\uz) = \frac{\Pi_k(\uz)}{\Pi_0(\uz)} = \frac{\log z_k}{2\pi i} + \ord(z_l)\,.
\eeq
Their inverse is the \emph{mirror map} $z_k(t) = z_k(t_1,t_2) = q_k + \ord(q_l^2)$. Inserting the mirror map into the holomorphic period $\Pi_0(\uz)$, we can write $\Pi_0$ as a holomorphic function of the $q_k$:
\beq
\Pi_0(\ut) := \Pi_0(\uz(\ut)) = 1 +\ord(q_k)\,.
\eeq

At this point, the form of the matrix $S$ is not entirely fixed, and we could define a new basis
\beq
\uPi(z) = \cR\uPi'(z)\,,\qquad \cR = \left(\begin{smallmatrix}1&0&0\\ 0&R&0\\0&0&1\end{smallmatrix}\right)\in\GL(4,\ZZ)\,.
\eeq
In this new basis the Gram matrix becomes
\beq
\label{eq:Sigma_prime}
\Sigma' = \cR^T\Sigma\cR = \left(\begin{smallmatrix} 0 & 0 & 1\\ 0&S'&0 \\ 1&0&0\end{smallmatrix}\right)\,,\textrm{~~with~~} S'= R^TSR\,.
\eeq
In ref.~\cite{lattice_paper} it is argued that, if we can find a basis such that the Gram matrix takes the form\footnote{In ref.~\cite{lattice_paper} the integer $n$ was restricted to be positive. This is immaterial, because we can always perform an additional rotation which changes the sign, e.g.,  $S = R= \left(\begin{smallmatrix}\phantom{-}0&-1\\-1&\phantom{-}0\end{smallmatrix}\right)$, then $R^TSR = -S$.}
\beq\label{eq:Sigma0}
\Sigma' = \left(\begin{smallmatrix}
0&\phantom{-}0&\phantom{-}0&\phantom{-}1\\
0&\phantom{-}0&\phantom{-}n&\phantom{-}0\\
0&\phantom{-}n&\phantom{-}0&\phantom{-}0\\
1&\phantom{-}0&\phantom{-}0&\phantom{-}0
\end{smallmatrix}\right)\,, \textrm{~~~for some integer $n\in \ZZ\setminus\{0\}$}\,,
\eeq
then the holomorphic period can be written as a product of two ordinary modular forms of weight 1, and the mirror map is a rational function of modular functions. Here we only summarise the main idea, and we refer to ref.~\cite{lattice_paper} for the complete argument. The transcendental lattice $\T(M)$ has signature $(2,2)$, and so the monodromy group $G_{\!M}$ is a subgroup of the orthogonal group $\OO(2,2)$. If $\SO_0(2,2)$ denotes the connected component of $\OO(2,2)$ containing the identity, then there is a well-known isomorphism
\beq
\SO_0(2,2) \simeq (\SL(2,\mathbb{R})\times\SL(2,\mathbb{R}))/\ZZ_2\,.
\eeq
The existence of this isomorphism is necessary, but not sufficient to guarantee that the periods and the mirror map admit a parametrisation in terms of ordinary modular forms. However, if the Gram matrix takes the form in eq.~\eqref{eq:Sigma_prime}, then a modular parametrisation always exists~\cite{lattice_paper}.
An example of a Feynman integral attached to a two-parameter family of K3 surfaces with this property was obtained in ref.~\cite{Dlapa:2024cje}, where it was observed that the K3 periods can be written as a product of complete elliptic integrals. In appendix~\ref{app:grav} we show how this results fits into the context of ref.~\cite{lattice_paper}. The purpose of this paper is to present two additional examples of Feynman integrals of this type, and to give a complete description of the K3 periods and the mirror map in terms of modular forms, functions and complete elliptic integrals.


\section{The conformal two-loop traintrack in two dimensions}
\label{sec:traintrack}

\subsection{The two-loop traintrack integral}

\begin{figure}[!t]
\centering
\begin{center}
\begin{tikzpicture}
    \node[draw, circle, fill=black, inner sep=1.5pt] (A) at (0,0) {};
    \node[draw, circle, fill=black, inner sep=1.5pt] (B) at (2,0) {};
    \node[draw, circle, fill=white, inner sep=1pt, label=above:$\alpha_5$] (C) at (1,1.5) {};
    \node[draw, circle, fill=white, inner sep=1pt] at (-0.74,0.54) {};
    \node[draw, circle, fill=white, inner sep=1pt] at (0,-1.05) {};
    \node[draw, circle, fill=white, inner sep=1pt] at (2.74,0.54) {};
    \node[draw, circle, fill=white, inner sep=1pt] at (2,-1.05) {};
    
    \draw[thick, black] (A) -- (B);
    \draw[thick, black] (B) -- (C);
    \draw[thick, black] (C) -- (A);
    
    \draw[thick] (A) -- (-0.7,0.5) node[left] {$\alpha_1$};
    \draw[thick] (A) -- (0,-1) node[left] {$\alpha_2$};
    \draw[thick] (B) -- (2.7,0.5) node[right] {$\alpha_3$};
    \draw[thick] (B) -- (2,-1) node[right] {$\alpha_4$};
\end{tikzpicture}
\end{center}
\caption{The conformal two-loop traintrack integral.}
\label{fig:traintrack}
\end{figure}
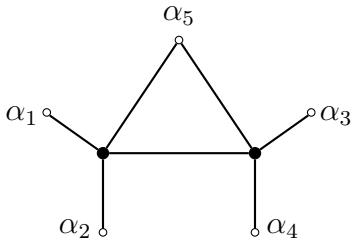

In this section we consider the Feynman integral (see figure~\ref{fig:traintrack}) 
\beq
\cI({\alpha}) = \int\frac{\rd^2\xi_1\,\rd^2\xi_2}{[(\xi_1,\alpha_1)(\xi_1,\alpha_2)(\xi_1,\alpha_5)(\xi_1,\xi_2)(\xi_2,\alpha_5)(\xi_2,\alpha_3)(\xi_2,\alpha_4)]^{1/2}}\,,
\eeq
where ${\alpha} = (\alpha_1,\ldots,\alpha_5)$ and $(\alpha,\beta) := (\alpha-\beta)^2$. 
This integral is conformally invariant with conformal weight $\frac{1}{2}$ at the external points $\alpha_i$, $i\le 4$, and conformal weight $1$ at the point $\alpha_5$, cf., e.g., refs.~\cite{Chicherin:2017frs,Loebbert:2020hxk}. In two dimensions it is convenient to introduce the complex variables
\beq
x_k = \xi_k^1+i\xi_k^2 \textrm{~~~and~~~}a_k = \alpha_k^1+i\alpha_k^2\,.
\eeq
The integral then takes the form
\beq
\cI({\alpha}) =\int\left(\prod_{k=1}^2\frac{\rd x_k\wedge \rd \overline{x}_k}{-2i}\right)\,\frac{1}{\sqrt{|P({x},{a})|^2}}\,,
\eeq
where we defined the polynomial
\beq
P({x},{a}) = (x_1-a_1)(x_1-a_2)(x_1-a_5)(x_2-x_1)(x_2-a_3)(x_2-a_4)(x_2-a_5)\,.
\eeq
Conformal invariance implies that we can write
\beq
\cI({\alpha}) = \big|F({a})\big|^2\,\phi(\uz)\,,
\eeq
where $F({a})$ is an algebraic function that carries the conformal weight and we introduced the conformal cross ratios $\uz=(z_1,z_2)$, with
\beq
z_1=\frac{a_{21}a_{45}}{a_{25}a_{41}} \textrm{~~~and~~~}z_2=\frac{a_{31}a_{45}}{a_{35}a_{41}}\,.
\eeq
There is some freedom in how to choose $F(\underline{a})$. Here we make the choice:
\beq
F(\underline{a}) = -\frac{1}{\sqrt{a_{14}\,a_{35}\,a_{52}}}\,,
\eeq
with $a_{ij}:=a_i-a_j$.

It follows from the discussion in refs.~\cite{Duhr:2022pch,Duhr:2023eld,Duhr:2024hjf} that the equation $y^2=P({x},{a})$ defines a two-parameter family of K3 surfaces, and the value of the Feynman integral can be related to the monodromy-invariant bilinear
\beq\label{eq:TT_phi}
\phi(\uz) = -\,\uPi(\uz)^\dagger\Sigma\uPi(\uz)\,,
\eeq
which computes the calibrated volume of the K3 surface, or the classical volume of the mirror family.\footnote{More generally, in ref.~\cite{Duhr:2022pch} it was argued that the Feynman integral computes the \emph{quantum} volume of the mirror. It is well known from string theory that for K3 surfaces, the quantum and classical volumes coincide.} In eq.~\eqref{eq:TT_phi}, $\uPi(\uz)$ is a vector of periods for the K3 surface (with the normalisation as in eq.~\eqref{eq:MUM-basis}), and $\Sigma$ is the corresponding intersection pairing. 
We choose the holomorphic differential
\beq
\Omega= \frac{1}{F({a})}\, \frac{\rd x_1\wedge \rd x_2}{\sqrt{P(x,{a})}}\,.
\eeq
Note that $\Omega$ is invariant under conformal transformations in one dimension.
The periods for this two-loop traintrack integral were evaluated in ref.~\cite{Duhr:2022pch}, in the special case $a_2=a_3$, in terms of complete elliptic integrals of the first kind, and in the case where no external points are identified, series expansions close to a MUM-point were given in refs.~\cite{Duhr:2022pch,Duhr:2024hjf}. In the remainder of this section we present closed solutions for the periods in terms of modular forms and complete elliptic integrals.

\subsection{The Appell $F_2$ system for the two-loop traintrack}
\label{sec:Appell_F2}

We start by using conformal invariance to fix $(a_1,a_4,a_5)$ to $(0,1,\infty)$. The holomorphic differential then takes the form
\beq
\Omega=-\frac{\rd x_1\wedge\rd x_2}{\sqrt{x_1(x_1-z_1)(x_2-x_1)(x_2-z_2)(x_2-1)}}\,.
\eeq
Letting $(x_1,x_2) = (z_1t_1,1-(1-z_2)t_2)$, we find
\beq
\Omega = \frac{\rd t_1\wedge\rd t_2}{\sqrt{t_1t_2(1-t_1)(1-t_2)(1-z_1t_1-(1-z_2)t_2)}}\,.
\eeq
Written in this form, $\Omega$ can be seen to define a family of Appell $F_2$ functions:
\begin{align}
\nonumber F_2(a;b,b'&;c,c';x,y) := \sum_{m,n=0}^{\infty}\frac{(a)_{m+n}(b)_m(b')_n}{(c)_m(c')_n}\,\frac{x^m}{m!}\,\frac{y^n}{n!}\\
&\,= \frac{\Gamma(c)\Gamma(c')}{\Gamma(b)\Gamma(b')\Gamma(c-b)\Gamma(c'-b')}\times\\
\nonumber&\,\times\int_0^1\rd t_1\,\rd t_2\,t_1^{b-1}\,t_2^{b'-1}\,(1-t_1)^{c-b-1}\,(1-t_2)^{c'-b'-1}\,(1-xt_1-yt_2)^{-a}\,,
\end{align}
where $(a)_n := \Gamma(a+n)/\Gamma(a)$ is the Pochhammer symbol. In particular, the holomorphic period is given by the Appell $F_2$ function
\beq
\Pi_0(\uz) = \pi^2\,F_2\big(\tfrac{1}{2};\tfrac{1}{2},\tfrac{1}{2};1,1;z_1,1-z_2)\,.
\eeq
The Appell $F_2$ function $F_2\big(\tfrac{1}{2};\tfrac{1}{2},\tfrac{1}{2};1,1;x,y)$ is a solution to the following system of partial differential equations:
\beq\label{eq:F2_system}
D_{xy}\omega = D_{yx}\omega= 0\,,
\eeq
where $D_{xy}$ is the differential operator
\begin{align}
    D_{xy}:=&\,x\,(1-x) \, {\partial_x^2 }-x\, y\, {\partial_x\partial_y }+(1-2 x)\, {\partial_x }-\frac{1}{2} \,y \,{\partial_y }-\frac{1}{4}\,.
\end{align}
It is well known that the solution space to the Appell $F_2$ differential system is four-dimensional, and the four solutions correspond to the four periods of our family of K3 surfaces. The two differential operators $D_{xy}$ and $D_{yx}$ can be taken as generators of the Picard-Fuchs differential ideal for our family of K3 surfaces.

We have solved the system in eq.~\eqref{eq:F2_system} locally close to $x=y=0$. Rescaling the variables such that $\xi=\frac{x}{16}$ and $\eta=\frac{y}{16}$ We find that a basis of solutions is given by
\begin{align}
\nonumber P_0(\xi,\eta)&\, = F_2\big(\tfrac{1}{2};\tfrac{1}{2},\tfrac{1}{2};1,1;16\xi,16\eta) \\
\nonumber&\, = 1+4\,\xi + 4\,\eta + 36\,\xi^2 + 48\,\xi\,\eta + 36\,\eta^2 + \ord(\xi^3)\,,\\
\nonumber P_1(\xi,\eta)&\, =P_0(\xi,\eta)\,\frac{\log (16\,\xi)}{2\pi i} + \frac{1}{2\pi i}\left(8\,\xi+8\,\eta +84\,\xi^2+128\,\xi\,\eta+96\,\eta^2+\ord(\xi^3)\right)\,,\\
\label{eq:P_basis}P_2(\xi,\eta)&\, =P_1(\eta,\xi)\,,\\
\nonumber P_3(\xi,\eta)&\, =P_0(\xi,\eta)\,\frac{\log (16\,\xi)}{2\pi i} \,\frac{\log (16\,\eta)}{2\pi i}\\
\nonumber&\,+\frac{\log (16\,\xi)}{(2\pi i)^2} \left(8\,\xi+8\,\eta+96\,\xi^2+128\,\xi\,\eta+84\eta^2+\ord(\xi^3)\right)\\
\nonumber&\,+\frac{\log (16\,\eta)}{(2\pi i)^2} \left(8\,\xi+8\,\eta+96\,\eta^2+128\,\xi\,\eta+84\,\xi^2+\ord(\xi^3)\right) + \\
\nonumber&\,+\frac{1}{(2\pi i)^2}\,\left(64\xi^2+128\,\xi\,\eta+ 64\eta^2+\ord(\xi^3)\right)\,.
\end{align}
From the structure of the solutions, we see that $\xi=\eta=0$ is a MUM-point. We define the canonical coordinates
\beq
t_k = \tau_k=\frac{P_k(\xi,\eta)}{2\pi i\, P_0(\xi,\eta)}\,.
\eeq
We can invert the previous relation, and we obtain the mirror map
\beq\bsp\label{eq:TT_mirror}
\xi(\tau_1,\tau_2) = & q_1 -8\,q_1^2 -8\,q_1\,q_2+ \ord(q_k^3)\,\\
\eta(\tau_1,\tau_2) = & q_2 -8\,q_2^2 -8\,q_1\,q_2 + \ord(q_k^3)\,,\qquad q_k = e^{2\pi i\tau_k}\,.
\esp\eeq
We can insert the mirror map into the holomorphic period to obtain:
\beq\bsp\label{eq:TT_holo}
P_0(\tau_1,\tau_2) &\,:= P_0(\xi(\tau_1,\tau_2) ,\eta(\tau_1,\tau_2) ) \\
&\,= 1+4\,q_1+4\,q_2+4\,q_1^2+4\,q_2^2-16\,q_1\,q_2+ \ord(q_k^3)\,.
\esp\eeq
We define the vector of solutions 
\beq
{P}(\xi,\eta) := (P_0(\xi,\eta),\ldots,P_3(\xi,\eta))^T\,.
\eeq
We find that it satisfies the quadratic relation:
\beq
{P}(\xi,\eta)^T\Sigma{P}(\xi,\eta) = 0\,, \textrm{~~~with~~~}\Sigma = \left(\begin{smallmatrix}0&\phantom{-}0&\phantom{-}0&\phantom{-}1\\
0&\phantom{-}0&-1&\phantom{-}0\\
0&-1&\phantom{-}0&\phantom{-}0\\
1&\phantom{-}0&\phantom{-}0&\phantom{-}0\end{smallmatrix}\right)\,.
\eeq
Hence, we are in a basis where the intersection pairing takes the form~\eqref{eq:Sigma0}, and so the holomorphic period and the mirror can be expressed in terms of modular forms and functions~\cite{lattice_paper}. 
In the remainder of this section, we will explicitly derive this modular representation of the mirror map and the holomorphic period in eqs.~\eqref{eq:TT_mirror} and~\eqref{eq:TT_holo}.

\subsection{The modular parametrisation}
Our strategy to find the representation of the holomorphic period and the mirror map in terms of modular forms follows two main steps. We first analyse the periods on the slice $\xi=\eta$, where we obtain a one-parameter family of K3 surfaces, whose periods admit a modular parametrisation. We then build an ansatz for a modular form in two parameters, and the free parameters in the ansatz for the holomorphic period are then fixed by symmetry and requiring that we recover the one-parameter result on the slice $\xi=\eta$.

\paragraph{The modular parametrisation on the slice $\xi=\eta$.} Let us start by analysing the one-parameter slice $\xi=\eta$. We obtain three independent solutions on this slice:
\begin{align}
\label{eq:Phat0}\widehat{P}_0(\xi) &\,:= P_0(\xi,\xi)\\
\nonumber&\,\phantom{:}=1+8\,\xi+120\, \xi^2+2240 \,\xi^3+37520\, \xi^4+\ord(\xi^5)\,,\\
\nonumber\widehat{P}_1(\xi) &\,:= P_1(\xi,\xi) = P_2(\xi,\xi)\,,\\
\nonumber\widehat{P}_2(\xi) &\,:= P_3(\xi,\xi)\,.
\end{align}
This reduction in the number of solutions is expected, because the dimension of the transcendental lattice is $2+m$, where $m$ is the number of independent moduli. The mirror map is given by
\beq\bsp
\tau &\,= \frac{\widehat{P}_1(\xi)}{2\pi i\,\widehat{P}_0(\xi)}\,,\\
\xi(\tau) &\,= q - 16 q^2 + 204 q^3 - 2432 q^4 + 27470 q^5 - 299712  q^6+\ord(q^7)\,.
\esp\eeq

It is known that the three periods of a one-parameter family are annihilated by a Picard-Fuchs operator of degree 3 which is the symmetric square of a Picard-Fuchs operator describing a family of elliptic curves. Hence, on the slice $\xi=\eta$, there must be a modular parametrisation for the periods.
In order to find this parametrisation, we start by analysing the holomorphic solution. We see that the sequence of Taylor coefficients in eq.~\eqref{eq:Phat0} matches precisely the sequence $s_{4B}(n)$ appearing in a Ramanujan-Sato series of level 4 (cf. ref.~\cite{Ramanujan-Sato}),
\beq
\widehat{P}_0(\xi) = \sum_{n=0}^\infty s_{4B}(n)\,\xi^n\,,
\eeq
with
\beq
s_{4B}(n) = \binom{2n}{n}\sum_{j=0}^n4^{n-2j}\binom{n}{2j}\binom{2j}{j}^2\,.
\eeq
It is known that Ramanujan-Sato series satisfy modular identities, which can be traced back to the existence of modular parametrisations. In this case we have~\cite{Ramanujan-Sato}
\beq
\xi(\tau) = \frac{1}{j_{4B}(\tau)+16}\,,
\eeq
with
\beq
j_{4B}(\tau) = \left(\frac{\eta(2\tau)}{\eta(4\tau)}\right)^{12} + 2^6\left(\frac{\eta(4\tau)}{\eta(2\tau)}\right)^{12}\,,
\eeq
where $j_{4B}(\tau)$ is a modular function of level 4 and $\eta(\tau)$ is the Dedekind eta function
\beq
\eta(\tau) = q^{1/24}\,\prod_{n=1}^\infty(1-q^n)\,,\qquad q=e^{2\pi i\tau}\,.
\eeq
The function $j_{4B}(\tau)$ corresponds to the entry A007247 in the On-Line Encyclopedia of Integer Sequences~\cite{sloane}, where it is shown how to express $j_{4B}(\tau)$ as a rational function in a Hauptmodul for $\Gamma_0(4)$.
Using that relation, we find
\beq
    \xi(\tau)=\frac{k(\tau)(1-k(\tau)^2)}{4\,(1-2k(\tau)-k(\tau)^2)^2}\,,
    \label{xtok}
\eeq
where $k(\tau)$ is a Hauptmodul for $\Gamma_0(4)$:
\beq\label{eq:k_def}
k(\tau) := \frac{\theta_2(\tau)^2}{\theta_3(\tau)^2}\,,
\eeq
and $\theta_i(\tau)$ are the standard Jacobi theta functions. This shows that the mirror  map of the one-parameter family admits a modular parametrisation in terms of modular functions of level 4. Consequently, we expect that the holomorphic period $\widehat{P}_0(\tau) := \widehat{P}_0(\xi(\tau))$ is an Eisenstein series of weight two and level four. The vector space of such Eisenstein series is finite, and a basis is easy to write down. By comparing $q$ expansions, we find
\beq\label{eq:1-param-traintrack}
\widehat{P}_0(\tau) := \theta_3(\tau)^4\,\big(1+2k(\tau) -k(\tau)^2\big)\,.
\eeq
Equations~\eqref{xtok} and~\eqref{eq:1-param-traintrack} provide the desired modular parametrisation on the slice $\xi=\eta$. Note that, close to the MUM-point $\xi=k=0$, we have the well-known relation
\beq\label{eq:theta3_to_K}
\theta_3(\tau)^2 = \frac{2}{\pi}\,\K\big(k(\tau)^2\big)\,,
\eeq
where $\K(\lambda)$ denotes the complete elliptic integral of the first kind,
\beq
\K(\lambda) = \int_0^1\frac{\rd t}{\sqrt{t(1-t)(1-\lambda t)}}\,.
\eeq
Hence, we can write the holomorphic period in terms of complete elliptic integrals,
\beq
\widehat{P}_0(\xi) = \frac{4}{\pi^2}\,\K(k^2)^2\,(1+2k -k^2)\,,
\eeq
where the relation between the variables $\xi$ and $k$ is given by the modular parametrisation of the mirror map in eq.~\eqref{xtok}.

\paragraph{The modular parametrisation for the Appell $F_2$ system.}
We now explain how we can leverage the knowledge of the modular parametrisation on the slice $\xi=\eta$ to a modular parametrisation on the whole moduli space. Since on the slice $\xi=\eta$ we needed modular functions of level four, we make the ansatz that $P_0(\tau_1,\tau_2)$ is a modular form of weight $(1,1)$ for $(\Gamma_1(4),\Gamma_1(4))$ (see appendix~\ref{app:modular}).\footnote{The congruence subroups $\Gamma_0(N)$ do not have modular forms of odd weight, while $\Gamma_1(N)\subseteq \Gamma_0(N)$ does for $N>2$. We therefore restrict ourselves to $\Gamma_1(4)$ to build our ansatz.}
This assumption restricts $P_0(\tau_1,\tau_2)$ to be of the form
\beq\label{eq:TT_ansatz}
P_0(\tau_1,\tau_2) = \theta_3(\tau_1)^2\,\theta_3(\tau_2)^2\,\rho(k_1,k_2)\,,
\eeq
where $\rho(k_1,k_2)=\rho(k_2,k_1)$ is a rational function of $k_i:=k(\tau_i)$. Since $P_0$ is holomorphic everywhere, $\rho$ cannot have poles, and so it must be a polynomial. The degree of this polynomial is bounded by the requirement of absence of poles at infinity. Using the fact that
\beq
\theta_3(\tau_i)^2 = \frac{2}{\pi}\,\K(k_i^2) = \ord\left(\tfrac{1}{k_i}\right)\textrm{~~~for } k_i\gg1\,,
\eeq
we see that $\rho(k_1,k_2)$ can at most be separately linear in both $k_1$ and $k_2$. Finally, the requirement that eq.~\eqref{eq:TT_ansatz} reduces to eq.~\eqref{eq:1-param-traintrack} on the slice $\xi=\eta$ fixes the form completely:
\beq\label{eq:P0Tok}
P_0(\tau_1,\tau_2) = \theta_3(\tau_1)^2\,\theta_3(\tau_2)^2\,\big(1+k(\tau_1)+k(\tau_2)-k(\tau_1)\,k(\tau_2)\big)\,.
\eeq
We have checked that this result reproduces the 25 first terms in the $q$-expansion of $P_0(\tau_1,\tau_2)$ obtained by direct computation (cf.~eq.~\eqref{eq:TT_holo}).

Similarly, for the mirror map $\xi(\tau_1,\tau_2) = \eta(\tau_2,\tau_1)$ we make the ansatz that it is  a rational function in $(k_1,k_2)$. In this case the functional form is not completely fixed by requiring that we recover the mirror map on the slice $\xi=\eta$ given in eq.~\eqref{xtok}. The remaining parameters are fixed by comparing to the $q$-expansion of the mirror map in eq.~\eqref{eq:TT_mirror}. We use the first few coefficients in eq.~\eqref{eq:TT_mirror} to fix all free coefficients, and we then automatically reproduce the 25 first terms in the expansion we have computed earlier. The result is
\beq\bsp\label{xtok2}
\xi(\tau_1,\tau_2) &\,=\frac{k_1\, (1-k_2^2)}{4 (1+k_1+k_2-k_1\, k_2)^2}\,,\\
    \eta(\tau_1,\tau_2)&\,= \frac{k_2\,(1-k_1^2)}{4 (1+k_1+k_2-k_1\, k_2)^2}\,.
    \esp\eeq
Equations~\eqref{eq:P0Tok} and~\eqref{xtok2} provide the modular parametrisation of the $F_2$ system. In the remainder of this section we explore some of its consequences. 

First, we can use eq.~\eqref{eq:theta3_to_K} to find a basis of solutions for the $F_2$ system in terms of complete elliptic integrals of the first kind,\footnote{The normalisation of this basis is slightly different from eq.~\eqref{eq:P_basis}. Since we are interested in presenting a basis only, this is immaterial.}
\beq\bsp\label{eq:PtoK_basis}
P_0(\xi,\eta) &\,= \frac{4}{\pi^2}\,(1+k_1+k_2-k_1\,k_2)\,\K(k_1^2)\,\K(k_2^2)\,,\\
P_1(\xi,\eta) &\,=  \frac{4}{\pi^2}\,(1+k_1+k_2-k_1\,k_2)\,\K(1-k_1^2)\,\K(k_2^2)\,,\\
P_2(\xi,\eta) &\,=  \frac{4}{\pi^2}\,(1+k_1+k_2-k_1\,k_2)\,\K(k_1^2)\,\K(1-k_2^2)\,,\\
P_3(\xi,\eta) &\,=  \frac{4}{\pi^2}(1+k_1+k_2-k_1\,k_2)\,\K(1-k_1^2)\,\K(1-k_2^2)\,.\\
\esp\eeq
The arguments of the elliptic integrals are algebraic functions of $(\xi,\eta)$,
\beq\bsp
k_1=& -\frac{\sqrt{\Sigma }}{8\,\xi},\\
k_2=& \frac{1}{2048\, \xi^2\, \eta (1-8 \,\xi-8 \,\eta)}\bigl\{\sqrt{\Lambda } (-1+16\, \xi +16\, \eta\\
&-128\, \xi^2-128\,\xi\,\eta+16\,\xi\,\sqrt{\Sigma}+16\,\eta\,\sqrt{\Sigma}+\sqrt{\Delta }-2 \sqrt{\Sigma })\\
&+2 \sqrt{\Sigma } (1-8\, \xi-8\, \eta) (1-16\, \eta-16\, \xi+128\, \xi\,\eta+\sqrt{\Delta })\bigr\},
\esp\eeq
where,
\beq\bsp
\Delta &=(1-16 \,\xi) (1-16\, \eta) (1-16 \,\xi-16\, \eta),\\
\Lambda &=2 (1-16\, \eta) (1-16\, \eta+\sqrt{\Delta })-32\, \xi (1-8\, \eta) (2-32\, \eta+\sqrt{\Delta })\\
&+128\, \xi^2 (5+128\, \eta^2-80\, \eta+\sqrt{\Delta })-2048\, \xi^3 (1-8\, \eta),\\
\Sigma &=1-16\, \eta-16\, \xi (1-8\, \eta)+64\, \xi^2 +\sqrt{\Delta }+\sqrt{\Lambda }.
\esp\eeq
In particular, this shows that the two-loop five-point traintrack integral in 2 dimensions can be written entirely in terms of complete elliptic integrals of the first kind and algebraic functions of the two cross ratios $(x,y) = (z_1,1-z_2)$.
Similarly, the canonical coordinates are given as ratios of elliptic integrals
\beq
\tau_a = i\,\frac{\K(1-k_a^2)}{\K(k_a^2)} = i\,\frac{P_a(\xi,\eta)}{P_0(\xi,\eta)}\,.
\eeq
We have checked that, if we change variables form $(\xi,\eta)$ to $(k_1,k_2)$ in the differential equation~\eqref{eq:F2_system}, then the four functions in eq.~\eqref{eq:PtoK_basis} solve the system. Note that this provides a rigorous proof that we have found the correct modular parametrisation. Moreover, the (almost) factorised form of the basis in eq.~\eqref{eq:PtoK_basis} allows us to write down a simpler set of differential operators that generate the Picard-Fuchs differential ideal. These differential operators are $\cD_{12}$ and $\cD_{21}$, with

\begin{align}
\nonumber\cD_{12} =&(1+k_1+k_2-k_1 k_2) [\left(k_1 k_2^3-3 k_1 k_2^2+k_1 k_2+k_1-k_2^3-3 k_2^2-k_2+1\right) \partial_{k_2}\\
\nonumber&+\left(k_1 k_2^4-k_1 k_2^3-k_1 k_2^2+k_1 k_2-k_2^4-k_2^3+k_2^2+k_2\right) \partial_{k_2}^2]\\
&-(1+4 k_1 k_2-k_2^2-k_1^2+k_1^2 k_2^2)\,,\\
\nonumber\cD_{21} =& \cD_{12}(k_1\leftrightarrow k_2)\,.
\end{align}

Second, we immediately see that the quadratic relations among periods are satisfied. For example, it is trivial to see that for the basis in eq.~\eqref{eq:PtoK_basis}
\beq
{P}(\xi,\eta)^T\Sigma{P}(\xi,\eta) = {P}(\xi,\eta)^T\Sigma\partial_{\xi}{P}(\xi,\eta) = {P}(\xi,\eta)^T\Sigma\partial_{\eta}{P}(\xi,\eta) = 0\,.
\eeq
Quadratic relations involving two derivatives are not necessarily zero. If we use the Legendre-type relation among elliptic integrals,
\beq
\K(1-\lambda)\,\partial_{\lambda}\K(\lambda) - \K(\lambda)\,\partial_{\lambda}\K(1-\lambda) = \frac{\pi}{4\lambda(1-\lambda)}\,,
\eeq
we immediately see that
\beq\bsp
{P}(\xi,\eta)^T&\Sigma_0\partial_{\xi}\partial_{\eta}{P}(\xi,\eta) = \frac{4}{\pi^2}\frac{1}{\xi\eta(1-16\xi-16\eta)}\\
&=\frac{64}{\pi^2}\,\frac{(1+k_1+k_2-k_1\,k_2)^6}{k_1\,k_2\,(1-k_1^2)\,(1-k_2^2)\,(1-k_1-k_2-k_1\,k_2)^2}\,.
\esp\eeq

Finally, let us comment on the modular properties of the solutions, which should be connected to the monodromy group of the periods. The monodromy group of the differential system in eq.~\eqref{eq:F2_system} was studied in ref.~\cite{F2_monodromy}, where it was found to be isomorphic to $\langle\iota_{12}\rangle\times \Gamma_{1,2}\times\Gamma_{1,2}$, where $\langle\iota_{12}\rangle$ is the $\mathbb{Z}_2$ group that exchanges $\tau_1$ and $\tau_2$, and $\Gamma_{1,2} = \Gamma_{\theta}/\{\pm\mathbbm{1}\}$, where $\Gamma_{\theta}$ is the theta-group, defined by
\beq\label{eq:Gamma_theta}
\Gamma_{\theta} = \Big\{\left(\begin{smallmatrix}a&b\\c&d\end{smallmatrix}\right)\in\SL(2,\mathbb{Z}) : ac = 0\!\!\!\!\mod 2 \textrm{~or~} bd = 0\!\!\!\!\mod 2\Big\}\,.
\eeq
We have the inclusion $\Gamma(2)\subset\Gamma_{\theta}$.
Clearly, $\det \iota_{12}=-1$, so if we restrict to the orientation-preserving part of the monodromy group, then we expect that that $P_0(\tau_1,\tau_2)$ is a modular form of weight $(1,1)$ in two variables for $(\Gamma_{1,2},\Gamma_{1,2})$.
Our results manifestly have the right transformation properties under $\Gamma_0(4)\subseteq \Gamma_{1,2}$, which is a subgroup of $\Gamma_{\theta}$. We have checked that the holomorphic solution and the mirror map define respectively a modular form of weight $(1,1)$ and a modular function for $\Gamma(2)\subseteq\Gamma_{1,2}$. However, we do not find that our basis has the correct transformation properties under $S = \left(\begin{smallmatrix}0&-1\\1&0\end{smallmatrix}\right) \in\Gamma_{\theta}$. 
This seems to be at odds with the findings of ref.~\cite{F2_monodromy}.
Since we have checked that our basis satisfies the $F_2$ differential system, we are sure that our result is correct. We do currently not have an explanation for this conundrum, and it would be interesting to investigate this further.


\section{Three-loop banana integrals with three equal masses}
\label{sec:bananas}

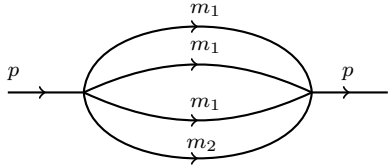
\begin{figure}[!h]
\centering
\begin{tikzpicture}
\coordinate (llinks) at (-2.5,0);
\coordinate (rrechts) at (2.5,0);
\coordinate (links) at (-1.5,0);
\coordinate (rechts) at (1.5,0);
\begin{scope}[very thick,decoration={
    markings,
    mark=at position 0.5 with {\arrow{>}}}
    ] 
\draw [-, thick,postaction={decorate}] (links) to [bend right=25]  (rechts);
\draw [-, thick,postaction={decorate}] (links) to [bend right=-25]  (rechts);

\draw [-, thick,postaction={decorate}] (links) to [bend left=85]  (rechts);
\draw [-, thick,postaction={decorate}] (llinks) to [bend right=0]  (links);
\draw [-, thick,postaction={decorate}] (rechts) to [bend right=0]  (rrechts);
\end{scope}
\begin{scope}[very thick,decoration={
    markings,
    mark=at position 0.5 with {\arrow{>}}}
    ]
\draw [-, thick,postaction={decorate}] (links) to  [bend right=85] (rechts);
\end{scope}
\node (d1) at (0,1.1) [font=\scriptsize, text width=.2 cm]{$m_1$};
\node (d2) at (0,0.6) [font=\scriptsize, text width=.2 cm]{$m_1$};
\node (d3) at (0,-0.15) [font=\scriptsize, text width=.2 cm]{$m_1$};
\node (d4) at (0,-.7) [font=\scriptsize, text width=0.3 cm]{$m_2$};
\node (p1) at (-2.0,.25) [font=\scriptsize, text width=1 cm]{$p$};
\node (p2) at (2.4,.25) [font=\scriptsize, text width=1 cm]{$p$};
\end{tikzpicture}
\caption{The three-loop banana graph with three equal masses.}
\label{fig:banana}
\end{figure}

Our second example will be the three-loop banana integrals in two dimensions (see figure~\ref{fig:banana}). 
When all four propagator masses are different, the maximal cuts compute the periods of a four-parameter family of K3 surfaces parametrised by $z_i=m_i^2/p^2$. In the following we are interested in the situation where three of the four masses are equal, i.e., we want to consider the integral 
\beq
\int\left(\prod_{j=1}^3\frac{\rd^Dk_j}{i\pi}\frac{1}{k_j^2-m_1^2}\right)\,\frac{1}{(p-k_1-k_2-k_3)^2-m_2^2}\,.
\eeq
In ref.~\cite{Bonisch:2020qmm} a set of generators for the Picard-Fuchs differential ideal was presented:
\begin{align}
\nonumber\cD_1=&  \left(-6 z_1 z_2-3 z_2^2+2 z_2\right) \theta_{z_1}^2+ \left(36 z_1^2+6 z_1 z_2-12 z_1-3 z_2^2+4 z_2-1\right) \theta_{z_2}^2\\
\nonumber&+\left(24 z_1 z_2+6 z_2^2-6 z_2\right) \theta_{z_1}\theta_{z_2}+ \left(3 z_2^2-z_2\right) \theta_{z_1}+ \left(30 z_1 z_2-3 z_2^2+z_2\right) \theta_{z_2}+ 6 z_1 z_2,\\
\cD_2=& \left(6 z_1 z_2+2 z_2^2-2 z_2\right) \theta_{z_1}^2+ \left(-24 z_1^2-2 z_1 z_2+8 z_1+2 z_2^2-2 z_2\right) \theta_{z_2}^2\\
\nonumber&+ \left(-12 z_1 z_2-4 z_2^2+4 z_2\right) \theta_{z_1}\theta_{z_2}+ \left(6 z_1 z_2-2 z_2^2\right) \theta_{z_1}+z_2 \left(2 z_2^2-14 z_1 z_2\right)\theta_{z_2}.
\end{align}
A Frobenius basis close to the MUM-point $z_i=0$  and the intersection pairing were given in refs.~\cite{Klemm:2019dbm,Bonisch:2020qmm} (see also ref.~\cite{Forum:2022lpz}). From this we can easily obtain a basis of local solutions for the case of three-equal masses, which we denote by 
\beq
\uPi(\uz) = \big(\Pi_0(\uz),\Pi_1(\uz),\Pi_2(\uz),\Pi_3(\uz)\big)^T\,.
\eeq
We normalise our basis as in eq.~\eqref{eq:MUM-basis}.
For the explicit form of the series expansions, we refer to refs.~\cite{Klemm:2019dbm,Bonisch:2020qmm}. 
Here it is sufficient to say that this basis satisfies the quadratic relations
\beq
\uPi(\uz)^T\Sigma_{31}\uPi(\uz)=0\,,
\eeq
with the intersection pairing\footnote{The intersection pairing in eq.~\eqref{eq:Sigma_31} differs from the one in ref.~\cite{lattice_paper} by the sign of the middle block. This can be traced back to a difference in the normalisation of the periods in eq.~\eqref{eq:MUM-basis}.}
\beq\label{eq:Sigma_31}
\Sigma_{31} = \left(\begin{smallmatrix}
 0 & 0 & 0 & 1 \\
 0 & -6 & -3 & 0 \\
 0 & -3 & 0 & 0 \\
 1 & 0 & 0 & 0 
 \end{smallmatrix}\right).
 \eeq
The canonical coordinates are defined as in eq.~\eqref{eq:t_def}.
Inverting this relation, we obtain the mirror map for the banana integral with three equal masses,
\beq\bsp\label{eq:mirror31}
z_1(t_1,t_2) &\,=q_1 +q_1^2+q_2 q_1-5 q_1^3-6 q_2 q_1^2
 + \ord(q_k^4)\,,\\
z_2(t_1,t_2) &\,= q_2+3 q_1 q_2-q_2^2-6 q_1^2 q_2-6 q_1 q_2^2+q_2^3
+\ord(q_k^4)\,,
\esp\eeq
with $q_k=e^{2\pi i t_k}$.
If we insert the mirror map into the series expansion of the holomorphic period, we find
\beq\bsp
\Pi_0(t_1,t_2) &\,:= \Pi_0(z_1(t_1,t_2),z_2(t_1,t_2))\\
&\,\phantom{:}=1+6 \left(q_1+q_2\right) q_1+ 36 q_2 q_1^3+6 \left(q_1^3+q_2^3\right) q_1^3+\ord(q_k^8)\,.
\esp\eeq
In ref.~\cite{lattice_paper}, it was shown that, if we define a new basis of periods according to
\beq
\label{eq:Pi_new}
\uPi^{\textrm{new}}(t_1,t_2) = M^{-1}\uPi(t_1,t_2)\,,\qquad \textrm{with } M =\left(\begin{smallmatrix}1&\phantom{-}0&\phantom{-}0&\phantom{-}0\\
0&\phantom{-}0&-1&\phantom{-}0\\
0&-1&\phantom{-}1&\phantom{-}0\\
0&\phantom{-}0&\phantom{-}0&\phantom{-}1
\end{smallmatrix}\right)\in \GL(4,\ZZ)\,,
\eeq
then in this basis the intersection matrix takes the form
\beq
M^T\Sigma_{31}M = \left(\begin{smallmatrix}
0&\phantom{-}0&\phantom{-}0&\phantom{-}1\\
0&\phantom{-}0&-3&\phantom{-}0\\
0&-3&\phantom{-}0&\phantom{-}0\\
1&\phantom{-}0&\phantom{-}0&\phantom{-}0
\end{smallmatrix}\right)\,.
\eeq
Hence, we know that it must be possible to express the periods ad the mirror map in terms of ordinary modular forms and functions. In the remainder of this section we derive this modular parametrisation. We proceed in the same way as for the traintrack integral in section~\ref{sec:traintrack}, and we use the equal-mass case $m_2=m_1$ as a boundary value. We therefore start by reviewing the equal-mass case.

\subsection{Review of the three-loop equal-mass banana integral}
The equal-mass case has been extensively studied in both the mathematics and physics literature~\cite{verrill1996,Bloch:2014qca,MR3780269,Primo:2017ipr,Klemm:2019dbm,Bonisch:2020qmm,Broedel:2019kmn,Broedel:2021zij,Pogel:2022yat}. Since in the equal-mass case we obtain a one-parameter family of K3 surfaces, the periods admit a modular parametrisation. We will not review the derivation of the modular parametrisation in detail, but we only summarise the result in a form that will be useful later on.

The maximal cuts of the equal-mass banana integral compute the periods of the one-parameter family of K3 surfaces studied in ref.~\cite{verrill1996}, and the holomorphic period is a modular form of weight 2 for $\Gamma_0(6)$. Written in our conventions, the holomorphic period takes the form:
\beq
\Pi_0^{\textrm{e.m.}}(\tau) = \frac{1}{12}\,(t(\tau)-3)^2\psi(\tau)^2\,,
\eeq
where $\psi(\tau)$ is a modular form of weight 1 for $\Gamma_1(6)$,
\beq\label{eq:em-banana-holo}
\psi(\tau) = \frac{2}{\sqrt{3}}\,\frac{\eta(3\tau)\,\eta(2\tau)^6}{\eta(\tau)^3\,\eta(6\tau)^2}\,,
\eeq
and $t(\tau)$ is a Hauptmodul for $\Gamma_0(6)$~\cite{Maier},
\beq
t(\tau) = 9\,\frac{\eta(6\tau)^8\,\eta(\tau)^4}{\eta(2\tau)^8\,\eta(3\tau)^4}\,.
\eeq
The mirror map is a modular function for $\Gamma_0(6)$, and hence a rational function in the Hauptmodul. One finds, with $z=m^2/p^2$,
\beq
z(\tau) = \frac{t(\tau)}{(t(\tau)-3)^2}\,.
\eeq
The holomorphic period  and the canonical coordinate can be written in terms of complete elliptic integrals of the first kind,
\beq\bsp\label{eq:tauToPsi}
\Pi_0^{\textrm{e.m.}}(z) &\, = \frac{1}{12}\,(t-3)^2\,\Psi_1(t)^2\,,\\
\tau &\,= -\frac{1}{2} + \frac{i\,\Psi_2(t)}{2\,\Psi_1(t)}\,,
\esp\eeq
where we defined~\cite{Laporta:2004rb}
\beq\bsp
\Psi_1(t)&\, = \frac{4 }{\pi  \sqrt[4]{(t-9) (t-1)^3}}\,\K\!\left(\Lambda(t)\right)\,,\\
\Psi_2(t)&\, = \frac{4 }{\pi  \sqrt[4]{(t-9) (t-1)^3}}\,\K\!\left(1-\Lambda(t)\right)\,,
\esp\eeq
with
\beq
\Lambda(t) = \frac{t^2-6 t+\sqrt{(t-9) (t-1)^3}-3}{2 \sqrt{(t-9) (t-1)^3}}\,.
\eeq


\subsection{The three-loop banana integral with three equal masses}

We now turn to the integral with three equal masses, and we start from the basis of periods in eq.~\eqref{eq:Pi_new}. It will turn out to be convenient to perform an additional rotation which rescales the periods,
\beq
\uPhi(\uz) := D^{-1}\uPi^{\textrm{new}}(\uz) = D^{-1}M^{-1}\uPi(\uz)\,,\qquad \textrm{with~}D = \diag\big(1,-\tfrac{1}{3},-1,1\big)\,.
\eeq
In this basis, the intersection pairing becomes
\beq
(MD)^T\Sigma_{31}(MD) = \left(\begin{smallmatrix}
0&\phantom{-}0&\phantom{-}0&\phantom{-}1\\
0&\phantom{-}0&-1&\phantom{-}0\\
0&-1&\phantom{-}0&\phantom{-}0\\
1&\phantom{-}0&\phantom{-}0&\phantom{-}0
\end{smallmatrix}\right)\,.
\eeq
Explicitly, the change of basis reads
\beq\bsp
\Phi_0(\uz) &\,= \Pi_0(\uz) \,,\\
\Phi_1(\uz) &\,= 3\,\Pi_1(\uz)  + 3\,\Pi_2(\uz)\,,\\
\Phi_2(\uz) &\,= \Pi_1(\uz)  \,,\\
\Phi_3(\uz) &\,= \Pi_3(\uz)  \,.
\esp\eeq
From this we can define a new set of canonical coordinates,
\beq\bsp 
\tau_1 &\,=3\,t_1+3\,t_2\,,\\
\tau_2 &\,=t_1\,.
\esp\eeq
If we insert this change of variables into the holomorphic period, we find
\beq\bsp
\Phi_0(\tau_1,\tau_2) &\,:= \Phi_0(z_1(\tau_1,\tau_2),z_2(\tau_1,\tau_2))\\
&\,\phantom{:}= 1+6 \left(\tilde{q}_1^2+\tilde{q}_2^2\right)+36 \tilde{q}_1^2 \tilde{q}_2^2+6 \left(\tilde{q}_1^6+\tilde{q}_2^6\right)+\ord(\tilde{q}_k^8)
\esp\eeq
with
\beq
\tilde{q}_1 = e^{2\pi i \tau_1/6} \textrm{~~~and~~~}\tilde{q}_2 = e^{2\pi i \tau_2}\,.
\eeq
Remarkably, we find that when expressed as a series in $(\tilde{q}_1,\tilde{q}_2)$, the holomorphic period becomes symmetric in $\tilde{q}_1$ and $\tilde{q}_2$:
\beq\label{eq:hidden_symmetry}
\Phi_0(6\tau_2,\tau_1/6) = \Phi_0(\tau_1,\tau_2)\,.
\eeq
We do currently not have a physics explanation for this hidden symmetry of the banana integral, which is not manifest at the level of the Feynman integral. If we express the mirror map in eq.~\eqref{eq:mirror31}, then $z_2$ has poles in $\tilde{q}_2$, but the poles cancel in the combination $z_1z_2$. We therefore propose that in the new basis, the natural complex structure moduli are
\beq\bsp\label{eq:xToZ}
Z_1(\tau_1,\tau_2) &\,:= x_1(\tau_1,\tau_2)  \\
&\phantom{:}=\tilde{q}_2+\tilde{q}_1^2+\tilde{q}_2^2-5 \tilde{q}_2^3-6 \tilde{q}_2 \tilde{q}_1^2
-7 \tilde{q}_1^4-6 \tilde{q}_2^2 \tilde{q}_1^2-7 \tilde{q}_2^4+\ord(\tilde{q}_k^5)\,,\\
Z_2(\tau_1,\tau_2) &\,:= x_1(\tau_1,\tau_2)\, x_2(\tau_1,\tau_2)\\
&\,\phantom{:}=\tilde{q}_1^2+4 \tilde{q}_2 \tilde{q}_1^2
-10 \tilde{q}_1^4-8 \tilde{q}_2^2 \tilde{q}_1^2+\ord(\tilde{q}_k^5)\,.
\esp\eeq

Having determined the holomorphic period and the mirror map in the new canonical coordinates, we can search for a modular parametrisation. We proceed in exactly the same manner as for the traintrack integral, and we make an ansatz
\beq
\Phi_0(\tau_1,\tau_2) = \psi\left(\tfrac{\tau_1}{6}\right)\,\psi(\tau_2)\,\rho_{31}(t_1,t_2)\,,
\eeq
where $\rho_{31}(t_1,t_2)$ is a rational function, and 
\beq
t_1 := t\left(\tfrac{\tau_1}{6}\right)\textrm{~~~and~~~}t_2 := t(\tau_2)\,.
\eeq
The hidden symmetry and the absence of poles require $\rho_{31}(t_1,t_2)$ to be a symmetric polynomial. Since $\psi(\tau) = \Psi_1(t)\sim \ord(1/t)$ for $t\gg 1$, this polynomial can be at most linear in both $t_1$ and $t_2$. Finally, the equal-mass limit corresponds to $\tau_1=6\tau_2$. Comparing to eq.~\eqref{eq:em-banana-holo}, we again find a unique solution:
\beq\label{eq:Phi031-modular}
\Phi_0(\tau_1,\tau_2) = \frac{1}{12}\,(t_1-3)\,(t_2-3)\,\psi\left(\tfrac{\tau_1}{6}\right)\,\psi(\tau_2)\,.
\eeq
We have checked that our ansatz reproduces the 25 first orders in the $q$-expansion of the holomorphic period. 

For the mirror map, however, unlike for the traintrack integral, we were not able to find an expression of $Z_1$ and $Z_2$ as rational functions in $t_1$ and $t_2$ (which are invariant under $\Gamma_0(6)$), but instead they can be written as rational functions in the Hauptmodul for $\Gamma_0(12)$,
\beq\bsp\label{eq:mirror31-modular}
Z_1(\tau_1,\tau_2) &\,= \frac{u_1^2-u_1 u_2^2+3 u_1+u_2^2}{\left(u_1^2+3\right) \left(u_2^2+3\right)},\\
Z_2(\tau_1,\tau_2) &\,= \frac{(u_1-3)^2 (u_1+1)^2 u_2^2}{\left(u_1^2+3\right)^2 \left(u_2^2+3\right)^2},
\esp\eeq
with
\beq
u_1 := u\left(\tfrac{\tau_1}{6}\right)\textrm{~~~and~~~}u_2 := u(\tau_2)\,,
\eeq
and $u(\tau)$ is a Hauptmodul for $\Gamma_0(12)$~\cite{Maier}:
\beq
u(\tau) := 3\,\frac{\eta(2\tau)^2\,\eta(12\tau)^4}{\eta(4\tau)^4\,\eta(6\tau)^2}\,.
\eeq
Note that the Hauptmodule for $\Gamma_0(6)$ and $\Gamma_0(12)$ are related by~\cite{Maier}
\beq\label{eq:tTou}
t(\tau) = \frac{u(\tau)\,(3-u(\tau))}{1+u(\tau)}\,.
\eeq

Equations~\eqref{eq:Phi031-modular} and~\eqref{eq:mirror31-modular} give the complete modular parametrisation of the maximal cuts for the three-loop banana integral with three different masses. The mirror map in eq.~\eqref{eq:mirror31-modular} is a modular function for $\Gamma_0(12)$, while the holomorphic period in eq.~\eqref{eq:Phi031-modular} is a modular form in two variables of weight (1,1) for $(\Gamma^1(6),\Gamma_1(6))$.\footnote{In appendix~\ref{app:modular}, we show that if $f(\tau)$ is a modular form of weight $k$ for $\Gamma_1(N)$, then $f(\tau/N)$ is a modular form of weight $k$ for $\Gamma^1(N)$.} Note that written in the form~\eqref{eq:Phi031-modular}, the hidden symmetry from~\eqref{eq:hidden_symmetry} has become completely manifest, and can be identified with exchanging the to copies of the periods of the sunrise elliptic curve.

The modular parametrisation allows us to write the holomorphic period and the canonical coordinates in terms of elliptic integrals. For the holomorphic period, we have
\beq\label{eq:Psi31}
\Phi_0(z_1,z_2) = \frac{1}{12}\,(t_1-3)(t_2-3)\,\Psi_1(t_1)\,\Psi_1(t_2)\,.
\eeq
Note that $t_1$ and $t_2$ are (very complicated) algebraic functions of $z_1$ and $z_2$, obtained by solving the sequence of equations in eqs.~\eqref{eq:xToZ},~\eqref{eq:mirror31-modular} and~\eqref{eq:tTou}. Likewise, the canonical coordinates take the form (cf.~eq.~\eqref{eq:tauToPsi})
\beq\bsp
\tau_1 &\,= -3 + \frac{3i\,\Psi_2(t_1)}{\Psi_1(t_1)}\,,\\
\tau_2 &\,= -\frac{1}{2} + \frac{i\,\Psi_2(t_2)}{2\,\Psi_1(t_2)}\,.
\esp\eeq
We have checked that, if we insert the sequence changes of variables into the differential equations satisfied by the maximal cuts of the banana integral, the holomorphic solution in eq.~\eqref{eq:Psi31} (as well as the logarithmically-divergent solutions obtained from it) satisfy those equations. Just like in the case of the traintrack integral, the (almost) factorised form of eq.~\eqref{eq:Psi31} allows us to find a simpler set of generators for the Picard-Fuchs ideal when we pass to the variables $(t_1,t_2)$. 



\section{Conclusion}
\label{sec:conclusions}

In this paper have presented two new examples of Feynman integrals related to a two-parameter of K3 surfaces whose periods can be written as a product of periods of families of elliptic curves. The existence of such a representation is guaranteed by the recent analysis in ref.~\cite{lattice_paper}, though this analysis provides no constructive way to identify the explicit modular parametrisation for the periods and the mirror maps. We have obtained these representations for the conformal traintrack and banana integral in two dimensions. Our strategy was to first restrict the two-parameter problem to a one-parameter slice, where we know that a modular parametrisation exists. We can then write down an ansatz for the two-parameter case, and fix the free coefficients in the ansatz by studying how it reduces to the one-parameter slice and by comparing coefficients in the expansion close to the MUM-point. Once we have fixed all free coefficients in our ansatz, we can check that our results satisfies the Picard-Fuchs differential equations. This last step provides a rigorous proof of our results.
We believe that the bootstrap approach that we have presented here can used to find modular parametrisations for other classes of K3 period. 

One of the main results our paper are explicit analytic results for the maximal cuts of the three-loop banana integral with three equal masses. Remarkably, the periods can be written as a product of two copies of the maximal cuts of the two-loop sunrise integral. We believe that our result may play an important role in determining a fully analytic result for this integral beyond the maximal cut. Indeed, the periods play an important role in determining a set of canonical differential equations. We expect that the full result for the three-loop banana integral with three equal masses can be expressed in terms of iterated integrals of modular forms, similarly to the existing results for the equal-mass two- and three-loop banana integrals. We leave this investigation for future work.

\subsection*{Acknowledgments}
We are grateful to Florian Loebbert for useful discussions.
This work was funded by the European Union (ERC Consolidator Grant LoCoMotive 101043686). Views and opinions expressed are however those of the author(s) only and do not necessarily reflect those of the European Union or the European Research Council. Neither the European Union nor the granting authority can be held responsible for them.

\appendix


\section{Modular forms for $\Gamma^1(N)$}
\label{app:modular}
In the following, $\Gamma$ will denote a subgroup of finite index\footnote{The index of a subgroup is the number of its cosets.} of the modular group $\SL(2,\mathbb{Z})$. 
A \emph{modular form} of weight $k$ for $\Gamma$ is a function $f:\mathbb{H}\to\mathbb{C}$ such that
\begin{enumerate}
\item $f$ is holomorphic on $\mathbb{H}$ and at the cusps $\mathbb{P}^1(\mathbb{Q}) = \mathbb{Q}\cup\{\infty\}$,
\item for all $\gamma=\left(\begin{smallmatrix}a&b\\c&d\end{smallmatrix}\right)\in\Gamma$, $f(\gamma\cdot \tau) = j(\gamma,\tau)^k\,f(\tau)$, with $j(\gamma,\tau) = c\tau+d$.
\end{enumerate}

The \emph{principle congruence subgroups of level $N$}, where $N>0$ is an integer, are defined by
\beq
\Gamma(N) = \big\{\gamma\in\SL(2,\mathbb{Z}):\gamma=\mathbbm{1}\!\!\mod N\big\}\,.
\eeq
In general, a congruence subgroup of level $N$ is a subgroup of $\SL(2,\mathbb{Z})$ that contains $\Gamma(N)$. Notable examples of congruence subgroups are
\beq\bsp
\Gamma_0(N) &\,= \Big\{\left(\begin{smallmatrix}a&b\\c&d\end{smallmatrix}\right)\in\SL(2,\mathbb{Z}) : c = 0\!\!\!\!\mod N\Big\}\,,\\
\Gamma_1(N) &\,= \Big\{\left(\begin{smallmatrix}a&b\\c&d\end{smallmatrix}\right)\in\SL(2,\mathbb{Z}) : c = 0\!\!\!\!\mod N\textrm{~and~}a,d = 1\!\!\!\!\mod N\Big\}\,,\\
\Gamma^0(N) &\,= \Big\{\left(\begin{smallmatrix}a&b\\c&d\end{smallmatrix}\right)\in\SL(2,\mathbb{Z}) : b = 0\!\!\!\!\mod N\Big\}\,,\\
\Gamma^1(N) &\,= \Big\{\left(\begin{smallmatrix}a&b\\c&d\end{smallmatrix}\right)\in\SL(2,\mathbb{Z}) : b = 0\!\!\!\!\mod N\textrm{~and~}a,d = 1\!\!\!\!\mod N\Big\}\,.
\esp\eeq
Another example of a congruence subgroup (of level 2) is the theta group from eq.~\eqref{eq:Gamma_theta}.
Note that we have the isomorphisms $\Gamma^0(N) \simeq \Gamma_0(N) $ and $\Gamma^1(N) \simeq \Gamma_1(N) $. The isomorphism is explicitly given by
\beq
\kappa_N\left(\begin{smallmatrix}a&b\\c&d\end{smallmatrix}\right) = \left(\begin{smallmatrix}a&Nb\\c/N&d\end{smallmatrix}\right)\,.
\eeq
If $f(\tau)$ is a modular form of weight $k$ for $\Gamma_0(N)$ or $\Gamma_1(N)$, then $f(\tau/N)$ is a modular form of weight $k$ for $\Gamma^0(N)$ or $\Gamma^1(N)$. Indeed, let $\gamma\in\Gamma^0(N)$. An easy computation shows that
\beq
\frac{1}{N}\,(\gamma\cdot \tau) = \kappa_N^{-1}(\gamma)\cdot \left(\frac{\tau}{N}\right)\,.
\eeq
Then
\beq\bsp
f\left(\frac{\gamma\cdot \tau}{N}\right) &\,= f\left(\kappa_N^{-1}(\gamma)\cdot \left(\frac{\tau}{N}\right)\right) \\
&\,= j\left(\kappa_N^{-1}(\gamma),\tfrac{\tau}{N}\right)^k\,f\left(\frac{\tau}{N}\right) \\
&\,= j(\gamma,\tau)^k\,f\left(\frac{\tau}{N}\right)\,.
\esp\eeq


\section{An example from gravitational wave physics}
\label{app:grav}

In ref.~\cite{Dlapa:2024cje} the computation of the universal (nonspinning) local-in-time conservative dynamics at fourth Post-Minkowskian order was presented. This computation requires the computation of Feynman integrals depending on two parameters $q$ and $x$. The precise definition is not relevant for the discussion here, and we refer to ref.~\cite{Dlapa:2024cje}. Here it suffices to note that the relevant Feynman integrals were computed using the method of differential equations~\cite{Kotikov:1990kg,Remiddi:1997ny,Gehrmann:1999as,Henn:2013pwa}. It was observed that the homogeneous solution of the differential equation involves the periods of a two-parameter family of K3 surfaces, and the periods can be written as products of complete elliptic integrals of the first type:
\beq\bsp
\pi^2\,\Pi_0(q,x) &\,= \K(-qx)\K\!\left(-\tfrac{q}{x}\right)\,,\\
\pi^2\,\Pi_1(q,x) &\,= \K(-qx)\K\!\left(1+\tfrac{q}{x}\right)\,,\\
\pi^2\,\Pi_2(q,x) &\,= \K(1+qx)\K\!\left(-\tfrac{q}{x}\right)\,,\\
\pi^2\,\Pi_3(q,x) &\,= \K(1+qx)\K\!\left(1+\tfrac{q}{x}\right)\,.
\esp\eeq
There is a MUM-point at $(q,x)=(0,0)$, and the vector of periods $\Pi(q,x) := \big(\Pi_0(q,x),\Pi_1(q,x),\Pi_2(q,x),\Pi_3(q,x)\big)^T$ satisfies the quadratic relation
\beq
\Pi(q,x)^T\Sigma\Pi(q,x) = 0\,,\qquad \Sigma = \left(\begin{smallmatrix}0&\phantom{-}0&\phantom{-}0&\phantom{-}1\\
0&\phantom{-}0&-1&\phantom{-}0\\
0&-1&\phantom{-}0&\phantom{-}0\\
1&\phantom{-}0&\phantom{-}0&\phantom{-}0
\end{smallmatrix}\right)\,.
\eeq
We see that the intersection form precisely takes the expected form. 

We can immediately write down the modular parametrisations for the periods. The mirror map is given by
\beq
q = -k(\tau_1)k(\tau_2)\textrm{~~~and~~~} x = \frac{k(\tau_1)}{k(\tau_2)}\,,
\eeq
where $k(\tau)$ is the Hauptmodul for $\Gamma_0(4)$ from eq.~\eqref{eq:k_def}, and we defined the canonical coordinates
\beq
\tau_1 = i\frac{\K(1+qx)}{\K(-qx)} \textrm{~~~and~~~} \tau_2 = i\frac{\K\!\left(1+\frac{q}{x}\right)}{\K\!\left(-\frac{q}{x}\right)}\,.
\eeq
The holomorphic period is then given by
\beq\bsp
\pi^2\,\Pi_0(q,x) &\,= \K(k_1)\K(k_2) = \frac{\pi^2}{4}\,\theta_3(\tau_1)^2\,\theta_3(\tau_2)^2\,.
\esp\eeq

\bibliographystyle{jhep}
\bibliography{twistedbib}

\end{document}